\documentclass[prd,nofootinbib,twocolumn,amsmath,amssymb,eqsecnum,longbibliography]{revtex4-1}
\usepackage[utf8]{inputenc}
\usepackage{mathtools}
\usepackage{graphics}
\usepackage{graphicx}
\usepackage{dcolumn}
\usepackage{bm}
\usepackage{dsfont} 
\usepackage{amsmath,amssymb}
\usepackage{hyperref}
\usepackage{tabularx}
\usepackage{epstopdf}
\usepackage{tensor}
\usepackage{mathrsfs}
\usepackage[normalem]{ulem}
\usepackage[usenames]{color}
\hypersetup{
    colorlinks=true,
    linkcolor=blue,
    filecolor=magenta,      
    urlcolor=blue,
    citecolor=blue
}
\usepackage[dvipsnames]{xcolor}
\usepackage{mathrsfs}

\newcommand{\inter}{\mathrm{(int)}}
\newcommand{\exter}{\mathrm{(ext)}}

\newcommand{\UVA}{Department of Physics, University of Virginia, P.O.~Box 400714, Charlottesville, VA 22904-4714, USA}

\AtBeginDocument{%
    \newwrite\bibnotes
    \def\bibnotesext{Notes.bib}
    \immediate\openout\bibnotes=\jobname\bibnotesext
    \immediate\write\bibnotes{@CONTROL{REVTEX41Control}}
    \immediate\write\bibnotes{@CONTROL{%
    apsrev41Control,author="08",editor="1",pages="1",title="0",year="1"}}
    \if@filesw
    \immediate\write\@auxout{\string\citation{apsrev41Control}}%
    \fi
}
\begin{document}

\title{Neutron Stars in Scalar-Tensor Theories: \\ Analytic Scalar Charges and Universal Relations}

 \author{Kent Yagi}
 \affiliation{\UVA}

 \author{Michael Stepniczka}
 \affiliation{\UVA}

\date{\today}

\begin{abstract} 

%NSs as tests of GR
Neutron stars are ideal astrophysical sources to probe general relativity due to their large compactnesses and strong gravitational fields.
%ST
For example, binary pulsar and gravitational wave observations have placed stringent bounds on certain scalar-tensor theories in which a massless scalar field is coupled to the metric through matter. 
%spontaneous scalar charges
A remarkable phenomenon of neutron stars in such scalar-tensor theories is spontaneous scalarization, where a normalized scalar charge remains order unity even if the matter-scalar coupling vanishes asymptotically far from the neutron star.
%Goal
While most works on scalarization of neutron stars focus on numerical analysis, this paper aims to derive accurate scalar charges \emph{analytically}. 
%What we do
To achieve this, we consider a simple energy density profile of the Tolman VII form and work in a weak-field expansion. We solve the modified Tolman-Oppenheimer-Volkoff equations order by order and apply Pad\'e resummation to account for higher order effects.
%What we find
We find that our analytic scalar charges in terms of the stellar compactness beautifully model those computed numerically.
%universal relation
We also find a quasi-universal relation between the scalar charge and stellar binding energy that is insensitive to the underlying equations of state. 
Comparison of analytic  scalar charges for Tolman VII  and constant density stars mathematically supports this quasi-universal relation. 
%punchline
The analytic results found here provide physically motivated, ready-to-use accurate expressions for scalar charges.

 \end{abstract}

\maketitle

\section{Introduction}

%NS obs, tests of nuclear physics, tests of GR
Neutron stars (NSs) are ideal compact astrophysical objects to probe fundamental physics. Due to their high central density that exceeds the saturation density of nuclear matter by several-fold, NSs  can efficiently test nuclear physics. For example, recent observations of $2M_\odot$ pulsars~\cite{Cromartie:2019kug}, x-ray emissions from hot spots on a rotating NS surface by NICER~\cite{Riley:2019yda,Miller:2019cac}, and gravitational waves from colliding NSs by LIGO and Virgo~\cite{TheLIGOScientific:2017qsa} have constrained properties of nuclear/quark matter and certain nuclear parameters (see e.g.~\cite{Abbott:2018exr,Annala:2017llu,Raithel:2018ncd,Lim:2018bkq,Bauswein:2017vtn,De:2018uhw,Most:2018hfd,Annala:2019puf,Malik2018,Zack:nuclearConstraints,Carson:2019xxz,Raithel:2019ejc,Chatziioannou:2020pqz,Raaijmakers:2019dks,Zimmerman:2020eho,Jiang:2019rcw,Dietrich:2020efo}). Due to their large compactnesses, NSs are also perfect sources to probe strong-field gravity. Indeed, observations of binary pulsars~\cite{Freire:2012mg,Yagi:2013qpa,Yagi:2013ava,Berti:2015itd,Shao:2017gwu,Archibald:2018oxs,Seymour:2019tir,Anderson:2019eay,Gupta:2021vdj} and the binary neutron star merger GW170817~\cite{Abbott:2018lct,Baker:2017hug,Sakstein:2017xjx,Ezquiaga:2017ekz,Zhao:2019suc,Silva:2020acr} have constrained various modifications to GR. 

%ST
Some of the most well-studied modified theories of gravity are scalar-tensor theories in which one introduces scalar fields (either minimally or non-minimally coupled to the metric) to the action. A simple scalar-tensor theory proposed by  Damour and Esposito-Far\`ese has two theoretical parameters $(\alpha_0,\beta_0)$ and allows NSs to scalarize \emph{spontaneously}~\cite{Damour:1992we,Damour:1993hw,Damour:1996ke}. Namely, when $\beta_0$ is sufficiently negative, NSs can have scalar charges of order unity even when $\alpha_0 \ll 1$ (or equivalently, when the value of the scalar field at infinity is small). Such a phenomenon can be understood as NSs undergoing a tachyonic instability, where the effective mass of the scalar field becomes imaginary~\cite{Ramazanoglu:2017xbl}. A catalog of NS scalar charges is provided in~\cite{Anderson:2019hio} while a surrogate model has recently been constructed in~\cite{Zhao:2019suc} based on \emph{numerical} calculations. The parameters $(\alpha_0,\beta_0)$ have been constrained with solar system experiments, various binary pulsars~\cite{Freire:2012mg,Shao:2017gwu,Archibald:2018oxs,Anderson:2019eay} and the binary NS merger event GW170817~\cite{Zhao:2019suc} to be $|\alpha_0| \lesssim 3 \times 10^{-4}$ and $\beta_0 \gtrsim -4.4$. Future forecasts on probing this scalar-tensor theory with black-hole/NS binaries have been made with pulsar~\cite{Berti:2015itd} and gravitational-wave~\cite{Shao:2017gwu,Carson:2019fxr,Zhao:2019suc} observations.

%Goal of this paper & what we do
The goal of this paper is to provide accurate expressions for scalar charges of NSs in the scalar-tensor theory in~\cite{Damour:1993hw} by solving the field equations \emph{analytically}. Such expressions are complementary to e.g. the surrogate model~\cite{Zhao:2019suc} mentioned earlier. To achieve this, instead of using realistic equations of state (EoSs) given in tables, we adopt the Tolman VII model~\cite{tolman} that approximates the energy density profile inside a NS to a quadratic function in a radial coordinate. Such a simple profile allows one to solve the Tolman-Oppenheimer-Volkoff (TOV) equations analytically in GR. Unfortunately, due to the complication of the modified TOV equations in scalar-tensor theories including the coupling with the scalar field, it is challenging to solve them analytically. To overcome this, we work in a weak-field approximation and solve the field equations order by order in the expansion\footnote{Similar calculations were carried out in quadratic gravity~\cite{Yagi:2015oca} and Einstein-\AE ther theory~\cite{Gupta:2021vdj} to respectively find  analytic scalar charges and sensitivities, which were then used in tandem with binary pulsar observations to place bounds on these theories.}. To take into account higher order contributions in the expansion, we apply Pad\'e resummation. We consider both spontaneous scalarization mentioned earlier and ``perturbative'' scalarization where the scalar charge is proportional to $\alpha_0$.

\begin{figure}[h]
%\begin{center}
\includegraphics[width=8.5cm]{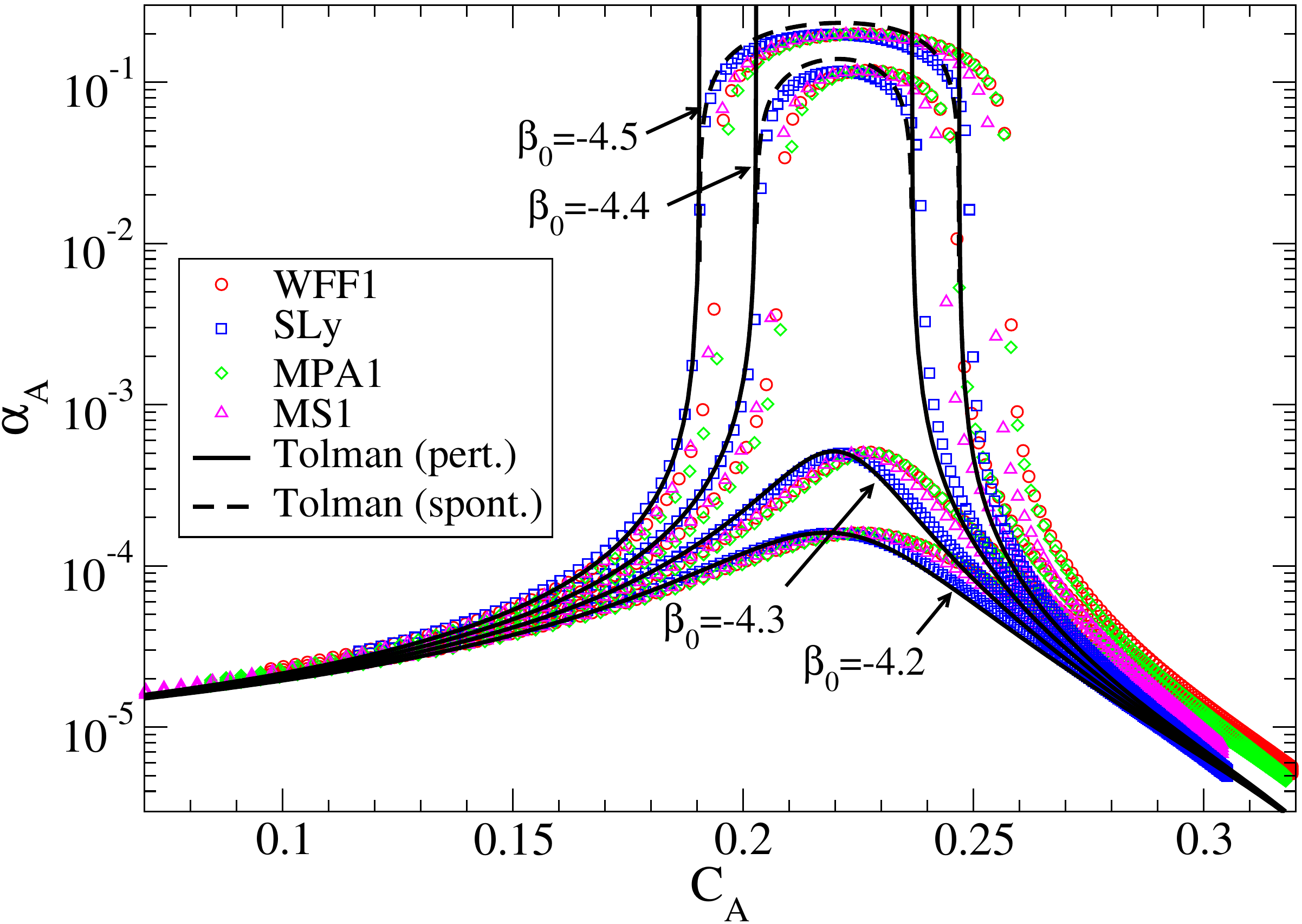}
\caption{\label{fig:charge-C-Tol} 
Comparison of analytic scalar charges with the Tolman VII model against numerical scalar charges with realistic EoSs as a function of the compactness. For the former, we show the results for both perturbative (solid) and spontaneous (dashed) scalarization. We present the results with four different choices of $\beta_0$. $\alpha_0$ is fixed to $\alpha_0=10^{-5}$ except for spontaneous scalarization for the Tolman case which assumes $\alpha_0=0$. Observe that the analytic Tolman results accurately model numerical results, especially with the SLy EoS. 
}
%\end{center}
\end{figure}

%What we find
Figure~\ref{fig:charge-C-Tol} summarizes our findings, which compares the analytic scalar charges with the Tolman VII model in terms of the stellar compactness (the ratio between the mass and radius) against those with realistic EoSs (WFF1~\cite{Wiringa:1988tp}, SLy~\cite{SLy}, MPA1~\cite{1987PhLB..199..469M}, MS1~\cite{Mueller:1996pm}) computed numerically. Notice that the former can accurately model the latter, especially for the SLy EoS, for both the spontaneous and perturbative scalarization cases. We also find that when we plot the scalar charges as a function of the stellar binding energy (the difference between the gravitational and baryonic mass), the relation between these quantities becomes quasi-universal and is insensitive to EoSs with an error of $\sim 1\%$ (see~\cite{Pappas:2018csu} for other universal relations involving scalar charges in scalar-tensor theories). We estimate analytically the amount of the quasi-universality by comparing analytic expressions for scalar charges with the Tolman VII model and with constant density stars and find that it is quasi-universal with an error of 0.3\%. The technique developed here should easily be applicable to other theories beyond GR to compute the stellar charges (or sensitivities).

%Organization and convention
The rest of the paper is organized as follows. In Sec.~\ref{sec:ST}, we review the scalar-tensor theory proposed in~\cite{Damour:1993hw}, explain the modified TOV equations, present the relation between mass and radius, and review spontaneous scalarization following~\cite{Horbatsch:2010hj}. In Sec.~\ref{sec:analytic}, we present the formalism for computing scalar charges analytically in the Tolman VII model by combining weak-field expansions and Pad\'e resummation. We compare such analytic results against numerical ones in Sec.~\ref{sec:results} and also present the quasi-universal relation between the scalar charge and binding energy. We conclude in Sec.~\ref{sec:conclusion} and give several different directions for future work. In Appendix~\ref{sec:CD}, we repeat the analytic calculation for constant density stars and derive scalar charges similar to the Tolman VII case.
We use the geometric units of $c=G=1$. The main expressions for scalar charges are summarized in a Supplemental Mathematica notebook~\cite{mathematica}.

\section{Scalar-tensor Theories and Neutron Stars}
\label{sec:ST}

We begin by reviewing scalar-tensor theories and spontaneous scalarization of NSs.

\subsection{Theory}

Presented here are the action and field equations for scalar-tensor theories in the Einstein frame. The former is given by~\cite{Damour:1992we,Damour:1993hw,Damour:1996ke}
\begin{equation}
S=\int d^{4} x \frac{\sqrt{-g}}{16\pi}\left(R-2 g^{\mu \nu} \partial_{\mu} \varphi \partial_{\nu} \varphi\right)+S_{\mathrm{mat}}\left[\psi, A^{2}(\varphi) g_{\mu \nu}\right]\,,
\end{equation}
where $R$ is the Ricci scalar for the metric $g_{\mu\nu}$ in this frame, $g$ is its determinant, $\varphi$ is the scalar field and $\psi$ is the matter field. $A$ is the conformal factor that relates $g_{\mu\nu}$ and the metric $\tilde g_{\mu\nu}$ in the (physical) Jordan frame as $\tilde g_{\mu\nu} = A(\varphi) g_{\mu\nu}$. Notice that we have set $G=1$ for the bare gravitational constant $G$.
Varying the above action with respect to $g_{\mu\nu}$ and $\varphi$, the field equations are given by~\cite{Damour:1993hw,Damour:1996ke}
\begin{eqnarray}
\label{eq:Ein_eq}
R_{\mu \nu}&=&2 \partial_{\mu} \varphi \partial_{\nu} \varphi+8\pi\left(T_{\mu \nu}-\frac{1}{2} g_{\mu \nu} T\right)\,, \\
\label{eq:scalar_eq}
\square \varphi&=&-4\pi \alpha(\varphi) T\,,
\end{eqnarray}
where $T_{\mu\nu}$ is the matter stress-energy tensor and
\begin{equation}
\alpha (\varphi) \equiv \frac{\partial \ln A(\varphi)}{\partial\varphi}\,. 
\end{equation}

In this paper, we consider an example scalar-tensor theory first considered by Damour and Esposito-Far\`ese~\cite{Damour:1992we,Damour:1993hw,Damour:1996ke} with
\begin{equation}
A(\varphi) = \exp\left( \frac{\beta_0}{2} \varphi^2 \right)\,.
\end{equation}
Another parameter of the theory is  $\alpha_0 \equiv \alpha(\varphi_0) = \beta_0 \varphi_0$ with $\varphi_0$ representing the scalar field at infinity.

\subsection{Neutron Stars}

To construct a NS solution, we begin with the metric ansatz given by
\begin{equation}
ds^2 = - e^{\nu(r)}dt^2 + e^{\lambda(r)}dr^2 + r^2 (d\theta^2 + \sin^2 \theta d\phi^2)\,.
\end{equation}
For the matter sector, we consider a perfect fluid whose stress-energy tensor in the Einstein frame is given by~\cite{Pani:2014jra}
\begin{equation}
T_{\mu\nu} = A^4(\varphi) \left[(\widetilde \rho + \widetilde P)u_\mu u_\nu + g_{\mu\nu}\widetilde P \right]\,,
\end{equation}
where $u^\mu$ is the 4-velocity of the fluid. Notice that $\widetilde \rho$ and $\widetilde P$ are the energy density and pressure in the Jordan frame that directly enters in the EoS $\widetilde P (\widetilde \rho)$. Plugging these into the field equations in Eqs.~\eqref{eq:Ein_eq} and~\eqref{eq:scalar_eq}, one finds~\cite{Damour:1993hw,Damour:1996ke}\footnote{There is also an equation for $\nu$ that we do not present here since it is unnecessary for deriving scalar charges.} 
\begin{eqnarray}
\label{eq:dmdr}
m^{\prime}&=&4 \pi r^{2} A^{4}(\varphi) \widetilde{\rho}+\frac{1}{2} r(r-2 m) \varphi'^{2}\,, \\
\label{eq:dPdr}
\widetilde{P}^{\prime}&=&-(\widetilde{\rho}+\widetilde{P})\left[\frac{m+4 \pi  A^4 \widetilde{P} r^3}{r(r-2 m)}+\frac{1}{2} r \varphi'^{2} +\alpha(\varphi) \varphi'\right]\,, \nonumber \\
\\
\label{eq:dphidrr}
\varphi'^{\prime}&=&4 \pi  \frac{r }{r-2 m}A^{4}(\varphi) \left[\alpha(\varphi)(\widetilde{\rho}-3 \widetilde{P})+r \varphi'(\widetilde{\rho}-\widetilde{P}) \right] \nonumber \\
&&-\frac{2(r-m)}{r(r-2 m)} \varphi'\,, 
\end{eqnarray}
where a prime represents an $r$ derivative and
\begin{equation}
e^{-\lambda(r)} = 1 - \frac{2m(r)}{r}\,.
\end{equation}

The asymptotic behavior of the scalar field at infinity is given by
\begin{equation}
\label{eq:varphi_asympt}
\varphi = \varphi_0 - \alpha_A\frac{M_A}{r} + \mathcal{O}\left(  \frac{M_A^2}{r^2}\right)\,,
\end{equation}
where 
\begin{equation}
\alpha_A = \frac{\partial \ln M_A}{\partial \varphi_0}
\end{equation}
is the scalar charge while $M_A$ is the total gravitational mass that can be read off from the asymptotic behavior of $m(r)$ at infinity as
\begin{equation}
\label{eq:m_inf}
m(r) = M_A + \mathcal{O}\left(  \frac{M_A}{r}\right)\,.
\end{equation}

\begin{figure}[h]
%\begin{center}
\includegraphics[width=8.5cm]{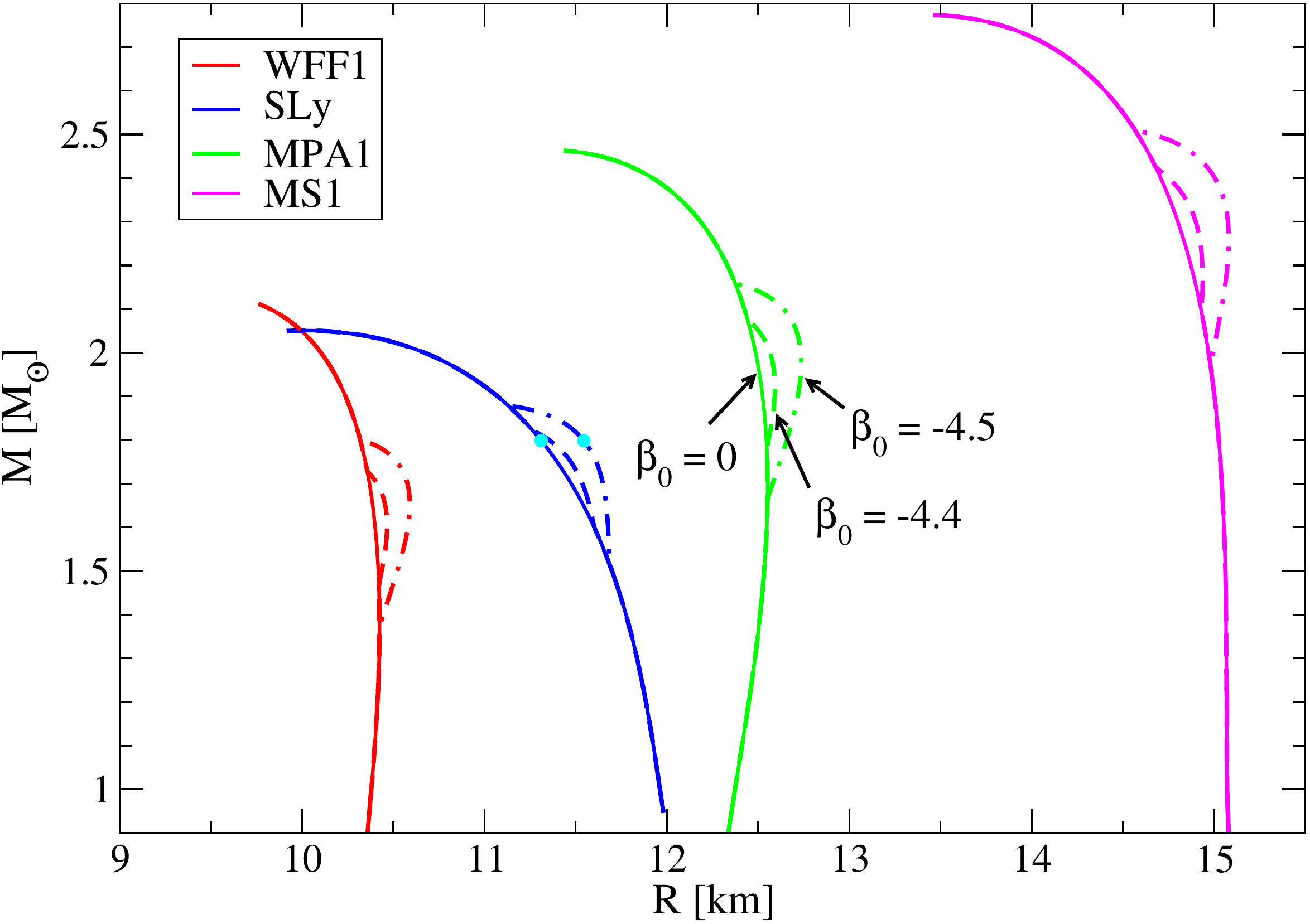}
\caption{\label{fig:MR} Mass-radius relation for NSs with four representative EoSs in GR ($\alpha_0=\beta_0=0$) and the scalar-tensor theory ($\alpha_0=10^{-5}$ and $\beta_0 = -4.4$, $-4.5$). The latter deviates from the former due to spontaneous scalarization. Two cyan dots correspond to the NSs of $1.8M_\odot$ with SLy EoS whose energy density profiles are shown in Fig.~\ref{fig:density_profile}.
}
%\end{center}
\end{figure}

We can construct NS solutions as follows. First, we impose the initial conditions of
\begin{equation}
m(0) = 0\,, \quad \widetilde P(0) = \widetilde P_c\,, \quad \varphi (0) = \varphi_c\,, \quad \varphi'(0) = 0\,,
\end{equation}
where $\widetilde P_c$ is the central pressure while $\varphi_c$ is the central value for the scalar field. Under these initial conditions together with a choice of $(\alpha_0,\beta_0)$ and an EoS, we solve Eqs.~\eqref{eq:dmdr}--\eqref{eq:dphidrr} in the interior region numerically and the stellar radius $R$ is determined by $\widetilde P(R) = 0$. We then solve the equations in the exterior region with $\widetilde P = \widetilde \rho = 0$. Finally, we read off $M_A$, $\varphi_0$ and $\alpha_A$ by comparing the asymptotic behavior of the numerically-computed $\varphi$ and $m$ with Eqs.~\eqref{eq:varphi_asympt} and~\eqref{eq:m_inf}. Figure~\ref{fig:MR} presents the mass-radius relation of NSs in GR and the scalar-tensor theory for four representative EoSs (WFF1~\cite{Wiringa:1988tp}, SLy~\cite{SLy}, MPA1~\cite{1987PhLB..199..469M}, MS1~\cite{Mueller:1996pm}) with different stiffness\footnote{These numerical results are computed with a Mathematica notebook developed in~\cite{Pani:2014jra}.}. ``Humps'' in the relation for the scalar-tensor theory correspond to NSs with spontaneous scalarization.

\subsection{Spontaneous Scalarization}
\label{sec:spnt}

An analytic attempt of computing scalar charges with spontaneous scalarization was taken in~\cite{Horbatsch:2010hj}. First, when $\alpha$ at the center of a star, $\alpha_c$, is small, one can expand both $\alpha_0$ and $\alpha_A$ in terms of $\alpha_c$ as\footnote{The scalar field (and its derivative) enter in even powers in Eqs.~\eqref{eq:dmdr} and~\eqref{eq:dPdr}, and in odd powers in Eq.~\eqref{eq:dphidrr}. This means that $\alpha_0$ enters in odd powers in $\alpha_c$ and $\alpha_A$. Inverting the former order by order (and substituting it to the latter), one finds that $\alpha_0$ and $\alpha_A$ enter in odd powers in $\alpha_c$.} 
\begin{eqnarray}
\label{eq:alpha0_in_alphac}
\alpha_0 &=& d_1 \alpha_c + d_2 \alpha_c^3 + \mathcal{O}(\alpha_c^5)\,, \\
\label{eq:alphaA_in_alphac}
\alpha_A &=& e_1 \alpha_c + e_2 \alpha_c^3 + \mathcal{O}(\alpha_c^5)\,.
\end{eqnarray}
Next, one can solve Eq.~\eqref{eq:alpha0_in_alphac} for $\alpha_c$ as
\begin{equation}
\label{eq:alpha_c}
\alpha_c = \omega C_+ + \bar \omega C_-\,,
\end{equation}
with 
\begin{equation}
\omega = 1, \quad -e^{\pm i \pi/3}\,,
\end{equation}
and
\begin{equation}
C_\pm = \left( \frac{\alpha_0}{2d_2} \pm \sqrt D \right)^{1/3}\,, \quad D = \left(\frac{\alpha_0}{2 d_{2}}\right)^{2}+\left(\frac{d_{1}}{3 d_{2}}\right)^{3}\,.
\end{equation}
From Eqs.~\eqref{eq:alphaA_in_alphac} and~\eqref{eq:alpha_c}, the scalar charge is given by
\begin{equation}
\alpha_A =  (\omega C_+ + \bar \omega C_-) e_1 +( \omega C_+ + \bar \omega C_-)^3 e_2  + \mathcal{O}(\alpha_c^5)\,.
\end{equation}
For example, when $\alpha_0 = 0$\footnote{Given current stringent bounds on $\alpha_0$, $\alpha_0 \approx 0$ is a valid approximation when discussing spontaneous scalarization.}, $\alpha_c$ can take non-vanishing values as $\alpha_c = \pm \sqrt{-d_1/d_2}$ and
\begin{equation}
\label{eq:charge_spnt}
\alpha_{A}=\pm\left[\left(-\frac{d_{1}}{d_{2}}\right)^{1 / 2} e_{1}+\left(-\frac{d_{1}}{d_{2}}\right)^{3 / 2} e_{2}\right]\,.
\end{equation}
Thus, although $\alpha_0 = 0$, the scalar charge becomes non-vanishing when $d_1/d_2<0$.

For constant density stars, $d_1$ and $e_1$ are given in a closed analytic form as~\cite{Horbatsch:2010hj}
\begin{eqnarray}
%\begin{aligned}
\label{eq:d1_CD}
d_{1} &=&\operatorname{HeunG}\left( \widetilde{a}, \widetilde{q} ; \widetilde{\alpha}_-, \widetilde{\alpha}_+, \frac{3}{2}, \frac{3}{2} ; Z \right) \nonumber \\
&&- \left(\widetilde Z \log \widetilde Z\right)  \, \operatorname{HeunG}'\left(\widetilde{a}, \widetilde{q} ; \widetilde{\alpha}_-, \widetilde{\alpha}_+, \frac{3}{2}, \frac{3}{2} ; Z\right)\,, \nonumber \\
\\
\label{eq:e1_CD}
e_{1} &=&\beta_0 \widetilde Z  \operatorname{HeunG}' \left(\widetilde{a}, \widetilde{q} ; \widetilde{\alpha}_-, \widetilde{\alpha}_+, \frac{3}{2}, \frac{3}{2} ; Z \right)\,,
%\end{aligned}
\end{eqnarray}
with 
\allowdisplaybreaks
\begin{eqnarray}
\widetilde{a}&=&-\frac{1}{1+3 \widetilde{p}_{c}}\,, \\
\widetilde{q}&=&\frac{3 \beta_0}{2}\left(\frac{3 \widetilde{p}_{c}-1}{3 \widetilde{p}_{c}+1}\right)\,, \\
\widetilde{\alpha}_\pm &= &\frac{3}{2}\left(1\pm \sqrt{1-\frac{8 \beta_0}{3}} \right)\,, \\
Z &=& \frac{\widetilde p_c}{1+3\widetilde p_c}\,, \\
\widetilde Z &= & \frac{1+\widetilde p_{c}}{1+3 \widetilde p_{c}}\,,
\end{eqnarray}
and $\widetilde p_c = \widetilde P_c/\widetilde \rho_c$ with $\widetilde P_c$ and $\widetilde \rho_c$ being the central pressure and energy density.  HeunG is the general Heun function (a generalization of the hypergeometric function) while its prime refers to a derivative with respect to $Z$. 

\begin{figure}[h]
%\begin{center}
\includegraphics[width=8.5cm]{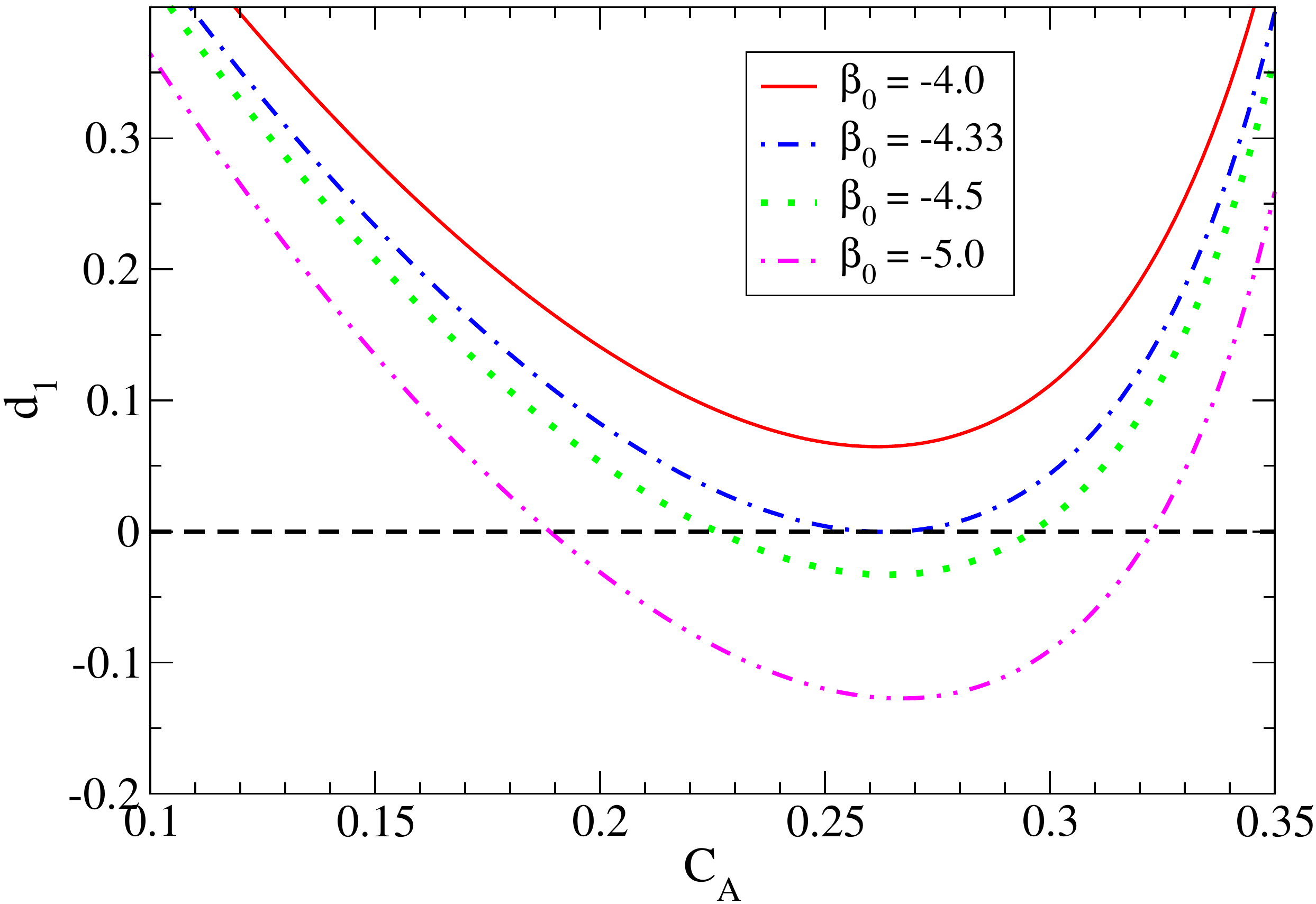}
\caption{\label{fig:d1} The linear coefficient $d_1$ in Eq.~\eqref{eq:alpha0_in_alphac} as a function of the stellar compactness $C_A$ for constant density stars with $\alpha_0=0$ and various choices of $\beta_0$. Spontaneous scalarization is realized when $d_1<0$.
}
%\end{center}
\end{figure}

Figure~\ref{fig:d1} shows $d_1$ for constant density stars as a function of the stellar compactness $C_A$ defined by
\begin{equation}
C_A = \frac{M_A}{R}\,.
\end{equation}
One can check numerically that spontaneous scalarization occurs when $d_1 < 0$~\cite{Horbatsch:2010hj}. This means that $d_2 > 0$ in the relevant parameter region, which we checked with the analytic expression for $d_2$ in Appendix~\ref{sec:CD}.
%For the relevant parameter region, $d_2 >0$ and thus the spontaneous scalarization can occur when $d_1 < 0$. 
Notice that 
 $\beta_0 = -4.33$~\footnote{See e.g.~\cite{Harada:1997mr} for a related, analytic estimate of the critical $\beta_0$ with constant density stars.
The critical value may vary slightly for neutron stars with other equations of state.} is the critical value below which the spontaneous scalarization is realized for a certain range of $C_A$. 
In~\cite{Horbatsch:2010hj}, closed analytic expressions are not found for $d_2$ and $e_2$.

\section{Analytic Scalar Charges}
\label{sec:analytic}

We now explain how we can use the weak-field expansion and Pad\'e resummation to analytically calculate NS scalar charges with the Tolman VII model.

\subsection{Formalism}

To construct approximate, analytic stellar solutions in the scalar-tensor theory, we work under a weak-field expansion. Namely, we decompose each unknown function $f(r)$ as 
\begin{equation}
\label{eq:wf}
f(r) = \sum_{k=0} f_k(r) \epsilon^k\,,
\end{equation}
where $\epsilon$ is a book-keeping parameter that counts the order of the GR compactness $C_0 = M_0/R$ where $M_0$ is the GR mass. We have
\begin{equation}
m_0(r) = \widetilde \rho_0(r) = \widetilde{p}_0(r) =   \widetilde{p}_1(r) = 0\,,
\end{equation}
and thus to leading order, $m(r) = \mathcal{O}(\epsilon)$, $\widetilde \rho(r) = \mathcal{O}(\epsilon)$ and $\widetilde p(r) = \mathcal{O}(\epsilon^2)$. Notice that $\widetilde \rho \gg \widetilde p$ in the weak-field limit. For the scalar field, we find it convenient to introduce 
\begin{equation}
\bar \varphi (r) \equiv \frac{\varphi (r)}{\alpha_0}\,,
\end{equation}
so that $\bar \varphi (r) = \mathcal{O}(\alpha_0^0)$ when $\alpha_0 \ll 1$. We decompose $\bar \varphi(r)$ as in Eq.~\eqref{eq:wf} and we have
\begin{equation}
\bar \varphi_0(r) = \frac{1}{\beta_0}\,.
\end{equation}

To see the effect of spontaneous scalarization mentioned in Sec.~\ref{sec:spnt}, we need to derive the scalar charge $\alpha_A$ valid up to $\mathcal{O}(\alpha_0^3)$ (which, in turn, means that it is valid to $\mathcal{O}(\alpha_c^3)$ as in Eq.~\eqref{eq:alpha0_in_alphac}). Namely, we need to find solutions up to to $\mathcal{O}(\alpha_0^2)$ higher than the leading. For this, we seek for a solution $\bar \varphi$ to $\mathcal{O}(\alpha_0^2)$. For $m(r)$, since $m_k = \mathcal{O}(\alpha_0^2)$ when $k \geq 2$ (namely, $m(r) = m_1(r)$ in GR), we need to find a solution for $m_k(r)$ valid to $\mathcal{O}(\alpha_0^4)$ such that $\alpha_A$ is valid up to $\mathcal{O}(\alpha_0^3)$. Since $\widetilde p$ does not enter in the $m'(r)$ equation in Eq.~\eqref{eq:dmdr}, we only need a solution for $\widetilde p$ to  $\mathcal{O}(\alpha_0^2)$. Next, we substitute the weak-field expansion of each unknown function in Eq.~\eqref{eq:wf} to the differential equations in Eqs.~\eqref{eq:dmdr}--\eqref{eq:dphidrr} and solve order by order in $\epsilon$. Below, we present the equations and solutions mainly to leading order in $\epsilon$ for an illustration purpose. In the actual calculation, we derived higher order contributions and applied a Pad\'e resummation that we explain later.

\subsection{Interior Solutions }

\begin{figure}[h]
%\begin{center}
\includegraphics[width=8.5cm]{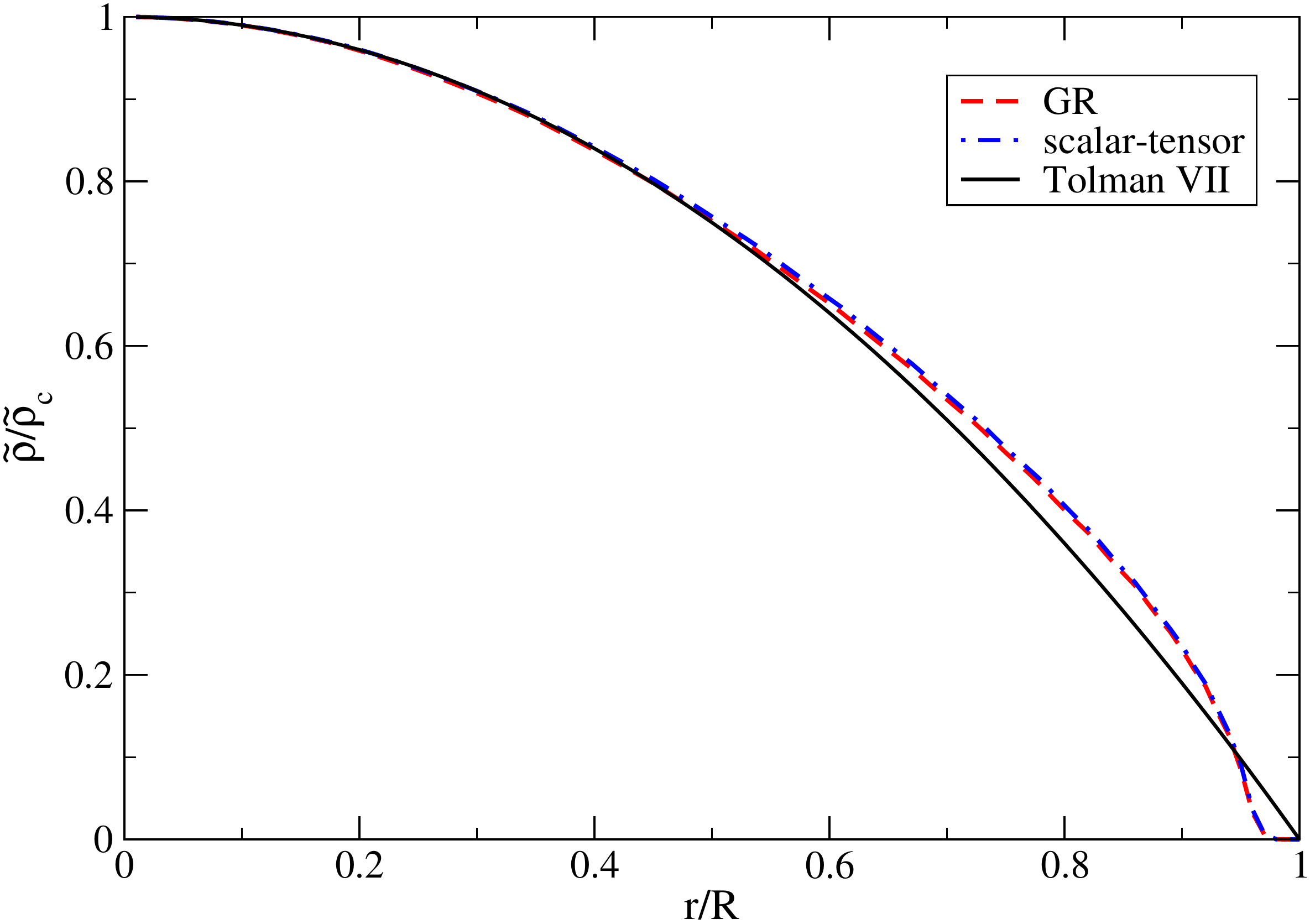}
\caption{\label{fig:density_profile} 
Energy density profiles (normalized by the central value) for NSs with $1.8M_\odot$ in GR and in the scalar-tensor theory with $\alpha_0=10^{-5}$ and $\beta_0 = -4.5$ (corresponding to cyan dots in Fig.~\ref{fig:MR}). We use the SLy EoS. For comparison, we also show the profile for the Tolman VII model given in Eq.~\eqref{eq:Tolman_density}. Observe that such a model can accurately describe the profile for realistic NSs.
}
%\end{center}
\end{figure} 

Since realistic EoSs are given in tables, one can only construct stellar solutions numerically. To overcome this, we use the Tolman VII model~\cite{tolman,Jiang:2019vmf} that is known to accurately model realistic NSs in GR. 

The density profile is given by a simple quadratic form as 
\begin{equation}
\label{eq:Tolman_density}
\widetilde \rho_1 =  \frac{15 M_0}{8 \pi 
   R^3} \left(1-\frac{r^2}{R^2}\right)\,, \quad \widetilde \rho_i (r) = 0 \quad (i\geq 2)\,.
\end{equation}
Thus, these stars are parameterized by $M_0$ and $R$, where the former is the stellar mass in GR. To check the validity of the above profile, we present in Fig.~\ref{fig:density_profile} the normalized energy density profile for NSs with $1.8M_\odot$ and the SLy EoS found numerically for both GR and scalar-tensor theory. For the latter, the star is spontaneously scalarized (see Fig.~\ref{fig:MR}). Notice that the energy density profiles in both theories are almost identical and can be approximated by Eq.~\eqref{eq:Tolman_density} shown by the black solid curve. This justifies the use of the Tolman VII model even for NSs in scalar-tensor theory. 

We next derive the field equations and solutions in the interior region. The leading differential equations are given by
\begin{eqnarray}
m_1'(r) &=&\frac{15}{2} \left( 1+ 4 \frac{\alpha_0^2}{\beta_0} +8 \frac{\alpha_0^4}{\beta_0^2}  \right)\frac{M_0}{R^5} r^2 (R^2-r^2)+ \mathcal{O}(\alpha_0^6)\,, \nonumber \\
\\
\bar \varphi_1''(r) &=&  -\frac{2 \bar \varphi_1'}{r} +\frac{15}{2} \left( 1+ 4 \frac{\alpha_0^2}{\beta_0} \right)\frac{M_0}{
   R^5}(R^2-r^2)+ \mathcal{O}(\alpha_0^4)\,, \nonumber \\
   \\
\widetilde P_2'(r) &=& -\frac{15 M_0}{8 \pi   R^5}(R^2-r^2)\left( \frac{m_1}{r^2} + \alpha_0^2 \bar\varphi_1'\right)\,. 
   \end{eqnarray}
Imposing the boundary conditions
\begin{equation}
\label{eq:BC1}
m_k(0)=0\,, \quad \bar \varphi_k(0)=\bar \varphi_{1c}\,, \quad \widetilde P_k(R)=0\,,
\end{equation}
we find
\begin{eqnarray}
m_1^\inter(r) &=& \frac{5}{2} \left( 1+ 4 \frac{\alpha_0^2}{\beta_0} +8 \frac{\alpha_0^4}{\beta_0^2}  \right)\frac{M_0}{R^3} r^3 \left( 1  - \frac{3}{5} \frac{r^2}{R^2} \right) \nonumber \\
   &&+ \mathcal{O}(\alpha_0^6)\,, \\
\bar \varphi_1^\inter(r) &=& \bar \varphi_{1c}+\frac{5}{4}  \left( 1+ 4 \frac{\alpha_0^2}{\beta_0} \right)\frac{M_0 }{
    R^3} r^2  \left( 1  - \frac{3}{10} \frac{r^2}{R^2} \right)\nonumber \\
   & &+ \mathcal{O}(\alpha_0^4)\,, \\
   \widetilde P_2^\inter(r) &=& \frac{15}{16 \pi  } \left( 1+  \frac{4+\beta_0 }{\beta_0} \alpha_0^2\right)\frac{M_0^2}{R^4} \nonumber \\
   &&\times\left(1-\frac{r^2}{R^2}\right)^2 \left(1-\frac{1}{2}\frac{r^2}{R^2}\right)+ \mathcal{O}(\alpha_0^4)\,. 
\end{eqnarray}

\subsection{Exterior Solutions and Perturbative Scalar Charges}

Next, we study the exterior region. By setting $\widetilde P = \widetilde \rho = 0$ in Eqs.~\eqref{eq:dmdr}--\eqref{eq:dphidrr}, we find
\begin{equation}
m_1'(r) = 0\,, \quad \bar \varphi_1''(r)=  -\frac{2 \bar \varphi_1'}{r}\,. 
\end{equation}
Imposing regularity at infinity, we can solve the above equations to find the exterior solution as\footnote{An analytic solution for a NS exterior in scalar-tensor theories without the weak-field expansion has been found in a different coordinate system~\cite{Damour:1992we,Damour:1996ke}. We found that working in the same coordinate system as the interior case is easier due to some nonlinearity in the coordinate transformation.}
\begin{equation}
\label{eq:m1_phi1_ext}
m_1^\exter(r) = M_1\,, \quad \bar \varphi_1^\exter(r) = \frac{\bar \omega_1}{r}\,,
\end{equation}
where $M_1$ and $\bar \omega_1$ are integration constants. 
We can determine the integration constants both in the interior and exterior solutions by imposing the boundary conditions at the surface:
\begin{eqnarray}
m_k^\inter(R) &=& m_k^\exter(R)\,, \\
\bar \varphi_k^\inter(R) &=& \bar \varphi_k^\exter(R)\,, \\
\bar \varphi_k'^\inter(R) &=& \bar \varphi_k'^\exter(R)\,. 
\end{eqnarray}
Similar to Eq.~\eqref{eq:m1_phi1_ext}, we can introduce integration constants $M_k$ and $\bar \omega_k$ at order $\epsilon^k$ as
\begin{eqnarray}
m_k^\exter(r) &=& M_k\left[1 + \mathcal{O}\left( \frac{M_0}{r} \right)\right]\,, \\ 
\bar \varphi_k^\exter(r) &=& \frac{\bar \omega_k}{r}+ \mathcal{O}\left( \frac{M_0^2}{r^2} \right)\,.
\end{eqnarray}

Once the integration constants are determined, we can compute the perturbative scalar charge from Eq.~\eqref{eq:varphi_asympt} as
\begin{equation}
\label{eq:charge}
\alpha_A 
=  - \alpha_0 \frac{\sum_{k=1}^N \bar \omega_k \epsilon^k}{
\sum_{k=1}^N M_k\epsilon^k}\,.
\end{equation}
In this paper, we computed up to $N=10$.
Equation~\eqref{eq:charge} allows us to express $\alpha_A$ in a series of the GR compactness $C_0$. It would be more useful to express the scalar charges in terms of the physical compactness in the scalar-tensor theory defined by 
\begin{equation}
C_A = \frac{M_A}{R}\,, \quad M_A = \sum_{k=1} M_k\epsilon^k\,.
\end{equation}
We can solve order by order to find $C_0$ in terms of $C_A$ and plug this into Eq.~\eqref{eq:charge}. To have the results consistent up to the order analyzed, we expand Eq.~\eqref{eq:charge} about $\alpha_0=0$ and $C_A=0$ and keep only to $\mathcal{O}(\alpha_0^3,C_A^{N})$. To have the series converge, we then construct a Pad\'e approximant of order $N/2$ (when $N$ is an even number) in $C_A$.

At leading order, the integration constants are derived as
\begin{eqnarray}
\bar \varphi_{1c} &=&- \frac{15}{8}  \left( 1+ 4 \frac{\alpha_0^2}{\beta_0} \right) \frac{M_0}{R}+ \mathcal{O}(\alpha_0^4)\,. \\
\label{eq:M1_Tol}
M_1 &=& \left( 1+ 4 \frac{\alpha_0^2}{\beta_0} +8 \frac{\alpha_0^4}{\beta_0^2}  \right) M_0+ \mathcal{O}(\alpha_0^6)\,, \\
\label{eq:omega1_Tol}
\bar \omega_1 &=& -\left( 1+ 4 \frac{\alpha_0^2}{\beta_0} \right) M_0+ \mathcal{O}(\alpha_0^4)\,.
\end{eqnarray}
\newpage
Using Eq.~\eqref{eq:charge}, the scalar charge to this order is given by 
\begin{eqnarray}
\label{eq:charge_leading}
\alpha_A &=& \alpha_0 \frac{\left( 1+ 4 \frac{\alpha_0^2}{\beta_0} \right)}{\left( 1+ 4 \frac{\alpha_0^2}{\beta_0} +8 \frac{\alpha_0^4}{\beta_0^2}  \right)} \nonumber \\
&=& \alpha_0 + \mathcal{O}(\alpha_0^5, C_A)\,.
\end{eqnarray}
We have derived $\alpha_A$ valid to $\mathcal{O}(C_A^{10})$. The first few terms are given by
\begin{eqnarray}
\alpha_A &= &\alpha_0-\frac{10}{7}
   \alpha_0 \left(\beta_0+1\right)  C_A+ \frac{5\alpha_0}{3003} \nonumber \\
  && \times \left(1253 \beta_0^2+1514
   \beta_0-1126\right)  C_A^2 + \mathcal{O}(\alpha_0^3, C_A^3)\,. \nonumber \\
\end{eqnarray}
We then construct a Pad\'e approximant to 5th order in $C_A$, which we provide in a Supplemental Mathematica notebook~\cite{mathematica}.

\subsection{Spontaneous Scalarization}

So far, we have constructed $\alpha_A$ in a series of $\alpha_0$ and $C_A$ in a form
\begin{equation}
\label{eq:alphaA_in_alpha0}
\alpha_A = \bar e_1(C_A) \alpha_0 + \bar e_3(C_A) \alpha_0^3 + \mathcal{O}(\alpha_0^5)\,,
\end{equation}
where $\bar e_1$ and $\bar e_3$ are functions of $C_A$ valid to $\mathcal{O}(C_A^N)$. We also have 
\begin{eqnarray}
\label{eq:alphac_in_alpha0}
\alpha_c &=& \alpha_0 \beta_0 \sum_{k=1}^N \bar \varphi_{kc} \epsilon^k \nonumber \\
&=& \bar d_1(C_A) \alpha_0 + \bar d_3(C_A) \alpha_0^3 + \mathcal{O}(\alpha_0^5)\,.
\end{eqnarray}
We can invert Eq.~\eqref{eq:alphac_in_alpha0} order by order in $\alpha_c$ to find $\alpha_0(\alpha_c)$ in a form given by Eq.~\eqref{eq:alpha0_in_alphac}. Then, we substitute this to Eq.~\eqref{eq:alphaA_in_alpha0} and expand about $\alpha_c=0$ to find $\alpha_A(\alpha_c)$ in a form in Eq.~\eqref{eq:alphaA_in_alphac}. Following Sec.~\ref{sec:spnt}, we can then find $\alpha_A$ for NSs under spontaneous scalarization.

In fact, it turns out that the contribution for the second term in Eq.~\eqref{eq:charge_spnt} is negligible and we focus on deriving $d_1$, $d_2$, and $e_1$.
We first find these coefficients in terms of a series expansion of $C_A$ valid to $\mathcal{O}(C_A^{10})$. The first few terms of these functions are given by 
\newline
\begin{widetext}
\begin{eqnarray}
d_1 &=& 1+\frac{15 \beta_0 }{8}C_A +\frac{5\beta_0}{128}  (17
   \beta_0-14)C_A^2+ \frac{\beta_0}{7168} \left(695 \beta_0^2-2294
   \beta_0-27720\right)
 C_A^3+ \mathcal{O}(C_A^4)\,,\\
d_2 & =&-\frac{1955
   \beta_0}{448} C_A^2+\frac{(1630328-3417469 \beta_0)
   \beta_0}{512512} C_A^3
+ \mathcal{O}(C_A^4)\,, \\
   e_1 &=& 1+\frac{5}{56} (5
   \beta_0-16) C_A
   +\frac{5}{384384}
   \left(5515 \beta_0^2-54170 \beta_0-144128\right) C_A^2 \nonumber \\
&&   + \frac{1}{156828672}\left(941985 \beta_0^3-19472682
   \beta_0^2-258360664 \beta_0-441114624\right) C_A^3+ \mathcal{O}(C_A^4)\,.
\end{eqnarray}
\end{widetext}
Next, we construct the Pad\'e approximants on these. We first tried 5th order Pad\'e approximants but found that there are some unphysical divergences (Fig.~\ref{fig:charge-C-Tol_conv}). Instead, we use 4th order Pad\'e approximants. We substituted these into the following expression to find scalar charges for NSs under spontaneous scalarization:
%\newpage
\begin{equation}
\label{eq:charge_spnt_simple}
\alpha_{A}=\pm\left(-\frac{d_{1}}{d_{2}}\right)^{1 / 2} e_{1}\,.
\end{equation}
The final expression is available in the Supplemental Mathematica notebook~\cite{mathematica}.

\section{Comparison with Numerical Results}
\label{sec:results}

Let us now compare the analytic scalar charge expression for the Tolman VII model with scalar charges for realistic NSs found numerically. Scalar charges can be both positive and negative but we focus on the former. 

\subsection{Scalar Charges vs Compactness}

\begin{figure}[h]
%\begin{center}
\includegraphics[width=8.5cm]{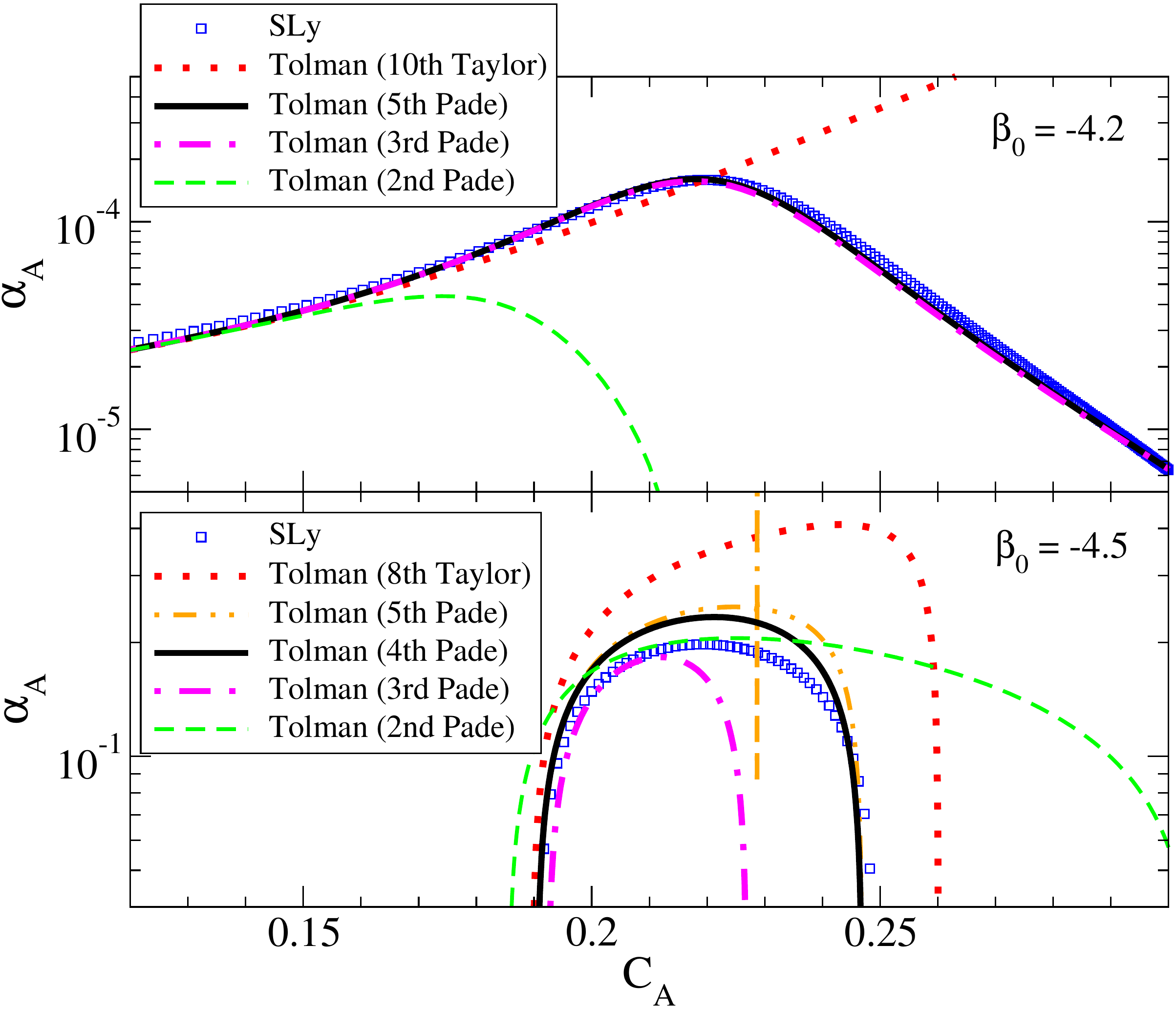}
\caption{\label{fig:charge-C-Tol_conv} Convergence of scalar charge calculations with the Tolman model for $(\alpha_0,\beta_0)=(10^{-5},-4.2)$  (top) and $(\alpha_0,\beta_0)=(0,-4.5)$ (bottom) under various approximations. We compare the Taylor series results and various Pad\'e resummation results against numerical results for the SLy EoS. Observe that Pad\'e resummation is crucial to accurately model the numerical result. The expressions for the black solid curves are the ones that we will use in the remaining part of this paper.}

%\end{center}
\end{figure} 

First, let us check the convergence of the analytic result for scalar charges in terms of the order of the Pad\'e resummation. Figure~\ref{fig:charge-C-Tol_conv} compares the analytic scalar charges at various orders of the Pad\'e approximants. We show the results for the perturbative scalarization case $(\alpha_0,\beta_0)=(10^{-5},-4.2)$ and the spontaneous scalarization case $(\alpha_0,\beta_0)=(0,-4.5)$. We also present Taylor series (without the Pad\'e resummation) up to $\mathcal{O}(C_A^{10})$ for perturbative scalarization and to $\mathcal{O}(C_A^{8})$ for spontaneous scalarization, as well as numerical results computed with the SLy EoS. Observe that for the perturbative scalarization case (top panel), the Pad\'e resummation at 3rd order is already a good approximation and is converging fast to the numerical result\footnote{The 3rd and 5th order Pad\'e results for the \emph{perturbative} scalar charge expressions have a noticeable difference for lower $\beta_0$ where the expressions diverge due to the onset of spontaneous scalarization (not shown in Fig.~\ref{fig:charge-C-Tol_conv}) and thus we use the 5th order result in the remainder of this section.}. On the other hand, the Taylor series result becomes less accurate when the compactness is relatively high. For the spontaneous scalarization case (bottom panel), there is little difference between the 4th and 5th order Pad\'e results, though the latter has some unphysical divergence. We therefore use the former to avoid this divergence. Additionally, the former has the advantages of a simpler functional expression and better agreement with the numerical SLy data.
%a clear difference in the Pad\'e expressions for spontaneous scalarization at 2nd, 3rd and 4th orders with the last one qualitatively agreeing with the numerical data.
All in all, these findings suggest that it is crucial to perform the resummation to find an accurate modeling.

Having understood the convergence, let us now carry out the comparison in more detail for various values of $\beta_0$ and EoSs. Figure~\ref{fig:charge-C-Tol} shows such a comparison. Observe that the analytic scalar charges for the Tolman VII case beautifully captures the numerical results for realistic NSs, especially for the SLy EoS. Notice that the agreement between the analytic and numerical results is good even for stars with spontaneous scalarization. This shows that the analytic result serves as an accurate, ready-to-use expression for the NS scalar charge in this scalar-tensor theory.

\subsection{Scalar Charges vs Binding Energy}

We next look at the relation between the scalar charge and the stellar binding energy, where the latter is defined by
\begin{eqnarray}
\Omega_A &=&-\frac{1}{2} \int d^{3} x \widetilde \rho(r) \int d^{3} x^{\prime} \frac{\widetilde \rho\left(r^{\prime}\right)}{\left|\boldsymbol{x}-\boldsymbol{x}^{\prime}\right|}\,, \\
&=& - 16 \pi^2\int_0^R dr  r^2 \widetilde \rho(r) \left(\int_r^R dr' r' \widetilde \rho(r')\right)\,,
\end{eqnarray}
where the second equality is valid only for spherically symmetric systems. Physically, this quantity measures the difference between the gravitational and baryonic mass of a star and can be used to e.g. probe certain formation scenario of NSs~\cite{Newton:2016weo}. For the Tolman VII model, the relation between the binding energy and compactness is given by
\begin{equation}
\label{eq:bind_comp_Tol}
\frac{\Omega_A}{M_A} = -\frac{5}{7}C_A + \mathcal{O}(\alpha_0^2)\,.
\end{equation}
Such a GR relation is sufficient to find the scalar charge expression in terms of the binding energy for perturbative scalar charges since the latter is already proportional to $\alpha_0$.
We can invert this relation and substitute it to the $\alpha_A(C_A)$ expressions. 

\begin{figure}[h]
%\begin{center}
\includegraphics[width=8.5cm]{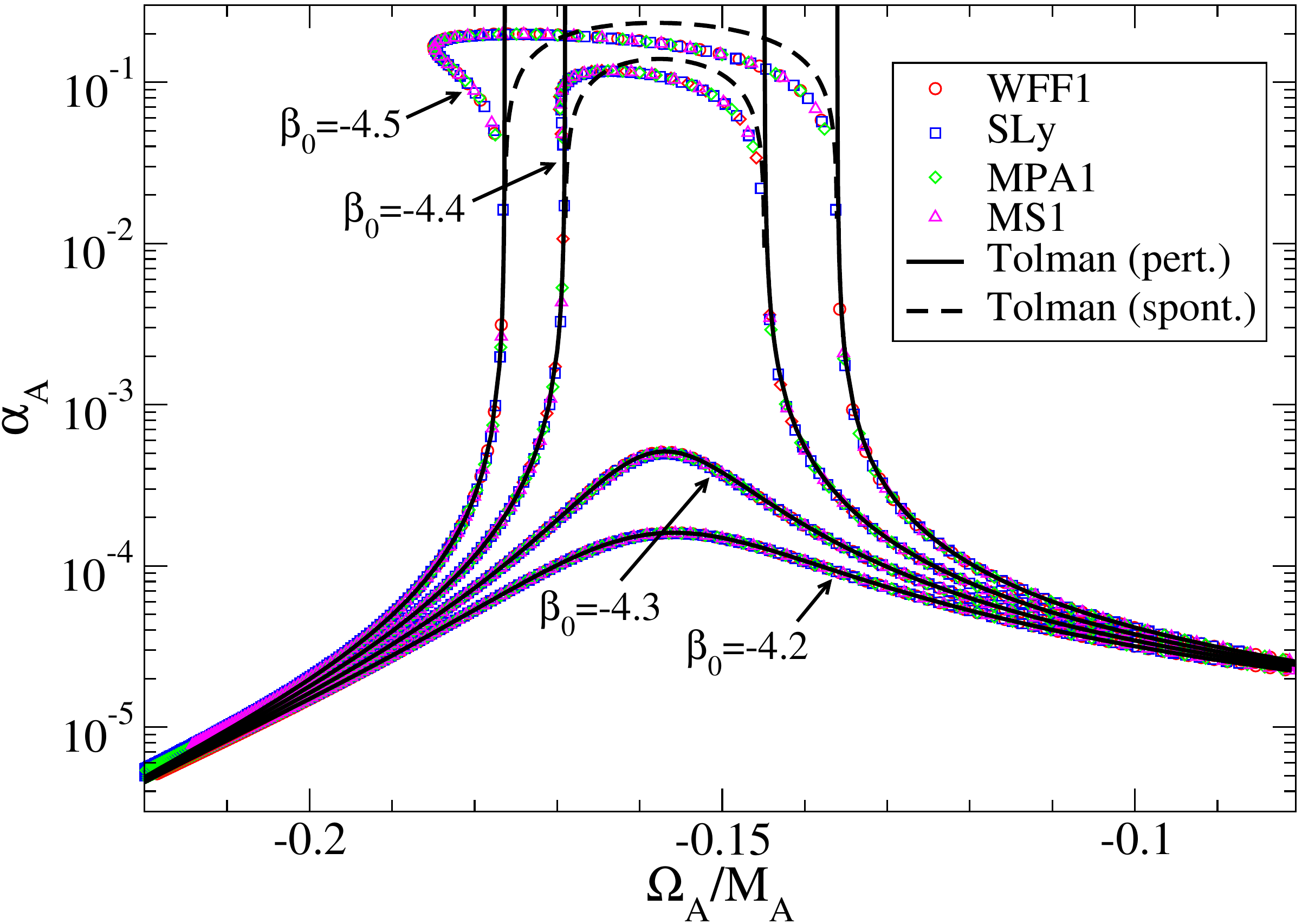}
\caption{\label{fig:charge-Bind-Tol} 
Similar to Fig.~\ref{fig:charge-C-Tol} but as a function of the binding energy. Notice that the results are quasi-universal and insensitive to the EoSs for fixed $\beta_0$.  
}
%\end{center}
\end{figure}

Figure~\ref{fig:charge-Bind-Tol} presents the scalar charge as a function of the binding energy (normalized by the stellar mass) for the same choices of $\alpha_0$ and $\beta_0$ and EoSs as in Fig.~\ref{fig:charge-C-Tol}. Observe  that numerical results are quasi-universal among different EoSs for a fixed $\beta_0$. Observe also that the analytic result accurately describes the numerical ones for perturbative scalarization. By taking the fractional difference between the numerical and analytic results for $\beta_0 = -4.2$ or $\beta_0 = -4.3$, we found that the quasi-universality holds to $\sim 1\%$ for $\Omega_A/M_A > -0.2$. 

\begin{figure}[h]
%\begin{center}
\includegraphics[width=8.5cm]{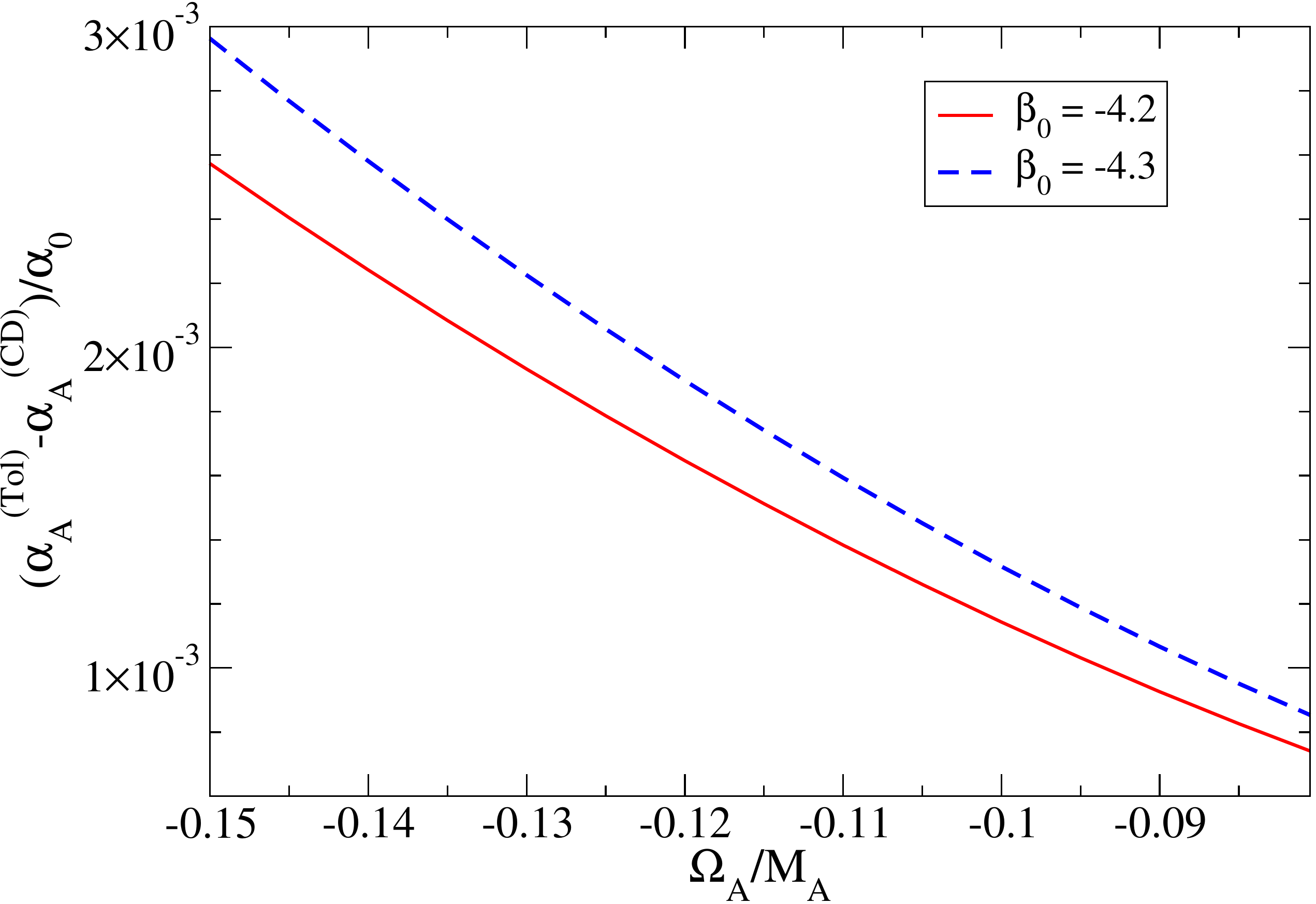}
\caption{\label{fig:charge-Bind-frac-diff} Fractional difference in the scalar charge between Tolman VII and constant density cases in terms of the binding energy (Eq.~\eqref{eq:alphaA_frac_diff}) for two choices of $\beta_0$. Notice that the two agree within an error of $\mathcal{O}(10^{-3})$, indicating the quasi-universality of the relation between the scalar charge and binding energy.
}
%\end{center}
\end{figure}

Analytic expressions help us to study the quasi-universal relation in more detail. When we expand the perturbative scalar charge expression for the Tolman VII model for small binding energy, we find 
\begin{eqnarray}
\label{eq:alpha_bind_Tol}
\alpha_A^{(\mathrm{Tol})} &= & \alpha_0 +2 \alpha_0  
   \left(\beta_0+1\right) \frac{\Omega_A}{M_A} \nonumber \\
&&   + \frac{7\alpha_0}{2145
  } \left(1253 \beta_0^2+1514
   \beta_0-1126\right)\frac{\Omega_A^2}{M_A^2} \nonumber \\
   &&+\mathcal{O}\left( \alpha_0^3, \frac{\Omega_A^3}{M_A^3} \right)\,. 
\end{eqnarray}
In Appendix~\ref{sec:CD}, we derive analytic scalar charges for constant density stars. We show that although $\alpha_A(C_A)$ is quite different from those for Tolman VII and realistic EoS cases, $\alpha_A(\Omega_A/M_A)$ is similar to the latter two cases, supporting the quasi-universal relation. For the perturbative scalar charge, the Taylor-series expansion in small binding energy is given by
\begin{eqnarray}
\alpha_A^{(\mathrm{CD})} &= & \alpha_0 +2 \alpha_0  
   \left(\beta_0+1\right) \frac{\Omega_A}{M_A} \nonumber \\
&&   + \frac{5\alpha_0}{21
   }  \left(17
   \beta_0^2+20 \beta_0-16\right)\frac{\Omega_A^2}{M_A^2} \nonumber \\
   && +\mathcal{O}\left( \alpha_0^3, \frac{\Omega_A^3}{M_A^3} \right)\,.
\end{eqnarray}
Comparing this with Eq.~\eqref{eq:alpha_bind_Tol}, we see that the two expressions are identical up to $\mathcal{O}(\Omega_A/M_A)$ and the difference only appears at $\mathcal{O}(\Omega_A^2/M_A^2)$. When taking the fractional difference between $\alpha_A^{(\mathrm{Tol})}$ and $\alpha_A^{(\mathrm{CD})}$, we find
\begin{eqnarray}
\label{eq:alphaA_frac_diff}
\frac{\alpha_A^{(\mathrm{Tol})} - \alpha_A^{(\mathrm{CD})}}{\alpha_0} &=& \frac{2 \left(311 \beta_0^2+1343
   \beta_0+1013\right)}{15015}\frac{\Omega_A^2}{M_A^2} \nonumber \\
  & & +\mathcal{O}\left( \alpha_0^3, \frac{\Omega_A^3}{M_A^3} \right)\,.
\end{eqnarray}
Figure~\ref{fig:charge-Bind-frac-diff} presents this fractional difference for $\beta_0 = -4.2$ and $-4.3$. Observe that the Tolman and constant density cases agree within an error of $\mathcal{O}(10^{-3})$. This provides strong analytic support for  the quasi-universality of the relation between $\alpha_A$ and $\Omega_A/M_A$.

For spontaneous scalarization, although the analytic expressions are valid as an order-of-magnitude estimate, the agreement with numerical results are not as good as the perturbative scalarization case. This is partially because we have used the relation between the binding energy and compactness in GR in Eq.~\eqref{eq:bind_comp_Tol}. Though, due to multi-valued scalar charges for some fixed binding energy with small $\beta_0$ (e.g. $\beta_0 = -4.5$), it would be difficult to find a closed, analytic expression for $\alpha_A(\Omega_A/M_A)$.

\section{Conclusion}
\label{sec:conclusion}

We derived analytically scalar charges for NSs in a scalar-tensor theory proposed in~\cite{Damour:1993hw}. This was achieved by considering the Tolman VII energy density profile. We first worked within the weak-field approximation and then resummed the series through Pad\'e approximants, deriving scalar charge expressions for both perturbative and spontaneous scalarization. We found that the analytic scalar charges are in excellent agreement with those computed numerically from realistic EoSs, especially for the SLy EoS. We also found a quasi-universal relation between the scalar charge and binding energy. The analytic expressions derived here allow one to mathematically support the quasi-universality by comparing the Tolman VII result with the constant density one. The analytic result provides an accurate, ready-to-use, and physically-motivated expression for scalar charges.

A similar quasi-universality between the stellar sensitivities and the binding energy were found recently in Einstein-\AE ther theory~\cite{Gupta:2021vdj}. The relation in the weak-field limit was first derived in~\cite{Foster:2007gr}. This was based on the result for weakly-gravitating stars in~\cite{Foster:2006az}, where the dipole moment for the vector field (which depends on the sensitivities) was derived within the parameterized post-Newtonian framework and all the internal structure dependence was found to be encoded in the binding energy. It would be interesting to study if a similar analysis can be carried out in scalar-tensor theories to explain the quasi-universal relation found here.

Various avenues exist for other possible future work. For example, one obvious extension is to apply the analysis presented here to other scalar-tensor theories with spontaneous scalarization, such as the ones with the mass potential~\cite{Chen:2015zmx,Ramazanoglu:2016kul,Yazadjiev:2016pcb,Doneva:2016xmf} or quartic interaction~\cite{Staykov:2018hhc}, scalar-Gauss-Bonnet gravity~\cite{Doneva:2017bvd,Silva:2017uqg,Doneva:2017duq,Silva:2018qhn,Minamitsuji:2018xde,Macedo:2019sem,Dima:2020yac,Doneva:2020nbb,Herdeiro:2020wei,Silva:2020omi}, or Horndeski theories~\cite{Andreou:2019ikc}. One may also apply the calculation to spontaneous vectorization~\cite{Ramazanoglu:2017xbl,Annulli:2019fzq,Kase:2020yhw}, tensorization~\cite{Ramazanoglu:2019gbz}, or spinorization~\cite{Ramazanoglu:2018hwk}. Another possibility might be to consider an analytic representation of the dynamical/induced scalarization~\cite{Barausse:2012da,Palenzuela:2013hsa,Shibata:2013pra,Sampson:2014qqa,Sennett:2016rwa,Khalil:2019wyy} in compact binary mergers. Finally, the parameter region of the coupling parameter $\beta_0$ that gives rise to spontaneous scalarization in the scalar-tensor theories studied here has been shown to be inconsistent with solar system experiments if one includes cosmological evolution of the scalar field~\cite{Damour:1992kf,Damour:1993id,Sampson:2014qqa,Anderson:2016aoi,Anderson:2017phb}. It would be interesting to consider analytic scalar charges in scalar-tensor theories that has a consistent cosmological evolution of the scalar field~\cite{Mendes:2014ufa,Mendes:2016fby,Antoniou:2020nax}.

\acknowledgments
K.Y. acknowledges support from NSF Grant PHY-1806776, NASA Grant 80NSSC20K0523, a Sloan Foundation Research Fellowship and the Owens Family Foundation. 
K.Y. would like to also acknowledge support by the COST Action GWverse CA16104 and JSPS KAKENHI Grants No. JP17H06358.

\appendix

\section{Scalar Charges for Constant Density Stars}
\label{sec:CD}

In this appendix, we repeat the calculations in the main text to find analytic expressions for scalar charges for constant density stars.
For such stars, we give the density profile as 
\begin{equation}
\widetilde \rho_1 =  \frac{3 M_0}{4 \pi  R^3}\,, \quad \widetilde \rho_i (r) = 0 \quad (i\geq 2)\,.
\end{equation}
Thus, these stars are parameterized by $M_0$ and $R$.
The leading differential equations are given by
\begin{eqnarray}
m_1'(r) &=&3 \left( 1+ 4 \frac{\alpha_0^2}{\beta_0} +8 \frac{\alpha_0^4}{\beta_0^2}  \right)\frac{M_0}{R^3} r^2+ \mathcal{O}(\alpha_0^6)\,, \\
\bar \varphi_1''(r) &=&  -\frac{2 \bar \varphi_1'}{r} +3 \left( 1+ 4 \frac{\alpha_0^2}{\beta_0} \right)\frac{M_0}{
   R^3}+ \mathcal{O}(\alpha_0^4)\,, \\
\widetilde P_2'(r)&=& -\frac{3 M_0}{4 \pi   R^3}\left( \frac{m_1}{r^2} + \alpha_0^2 \bar\varphi_1'\right)\,.
   \end{eqnarray}
The last equation is valid to full order in $\alpha_0$. Imposing the boundary condition as in Eq.~\eqref{eq:BC1},
one can solve the above differential equations in the interior region to find
\begin{eqnarray}
m_1^\inter(r) &=& \left( 1+ 4 \frac{\alpha_0^2}{\beta_0} +8 \frac{\alpha_0^4}{\beta_0^2}  \right)\frac{M_0}{R^3} r^3+ \mathcal{O}(\alpha_0^6)\,, \\
\bar \varphi_1^\inter(r) &=& \bar \varphi_{1c}+\frac{1}{2}  \left( 1+ 4 \frac{\alpha_0^2}{\beta_0} \right)\frac{M_0 }{
    R^3} r^2+ \mathcal{O}(\alpha_0^4)\,, \\
   \widetilde P_2^\inter(r) &=& \frac{3}{8 \pi  } \left( 1+  \frac{4+\beta_0 }{\beta_0} \alpha_0^2\right)\frac{M_0^2}{R^4} \left(1-\frac{r^2}{R^2}\right)+ \mathcal{O}(\alpha_0^4)\,. \nonumber \\
\end{eqnarray}

Next, we study the exterior solution and the scalar charge. At leading order in the weak-field expansion, the integration constants are determined as
\begin{eqnarray}
\bar \varphi_{1c} &=&- \frac{3}{2}  \left( 1+ 4 \frac{\alpha_0^2}{\beta_0} \right) \frac{M_0}{R}+ \mathcal{O}(\alpha_0^4)\,. \\
M_1 &=& \left( 1+ 4 \frac{\alpha_0^2}{\beta_0} +8 \frac{\alpha_0^4}{\beta_0^2}  \right) M_0+ \mathcal{O}(\alpha_0^6)\,, \\
\bar \omega_1 & =& -\left( 1+ 4 \frac{\alpha_0^2}{\beta_0} \right) M_0+ \mathcal{O}(\alpha_0^4)\,.
\end{eqnarray}
Notice that $M_1$ and $\bar \omega_1$ are the same as the Tolman VII  case as in Eqs.~\eqref{eq:M1_Tol} and~\eqref{eq:omega1_Tol}. Thus, the scalar charge is also same as in Eq.~\eqref{eq:charge_leading}.
Similar to the Tolman case, we then construct a Pad\'e approximant to 5th order in $C_A$ that we provide in the Supplemental Mathematica notebook~\cite{mathematica}.

Let us now find the expression for spontaneous scalarization. Although $d_1$ and $e_1$ are fully given in Eqs.~\eqref{eq:d1_CD} and~\eqref{eq:e1_CD}, we derived  $d_1$ and $e_1$ in a 4th order Pad\'e approximant form to make the scalar charge expression similar to that for the Tolman case discussed in the main text. 
We have carried out a similar analysis for $d_2$  whose full expression has not been found yet. The first few terms are given by
%\begin{widetext}
\begin{align}
d_2 =&-\frac{21 \beta_0
  }{8}C_A^2-\frac{3}{560} \beta_0 (547 \beta_0-278) C_A^3 + \mathcal{O}(C_A^4)\,.
\end{align}
%\end{widetext}
We have derived $d_2$ to $\mathcal{O}(C_A^{8})$ and constructed a 4th order Pad\'e approximant. 
We can substitute these Pad\'e resummed forms for $d_1$, $d_2$, and $e_1$  to Eq.~\eqref{eq:charge_spnt_simple} to find the scalar charge for constant density stars under spontaneous scalarization when $\alpha_0 \ll 1$. We provide the final expression in the Supplemental Mathematica notebook~\cite{mathematica}.

\begin{figure}[!htbp]
\includegraphics[width=8.5cm]{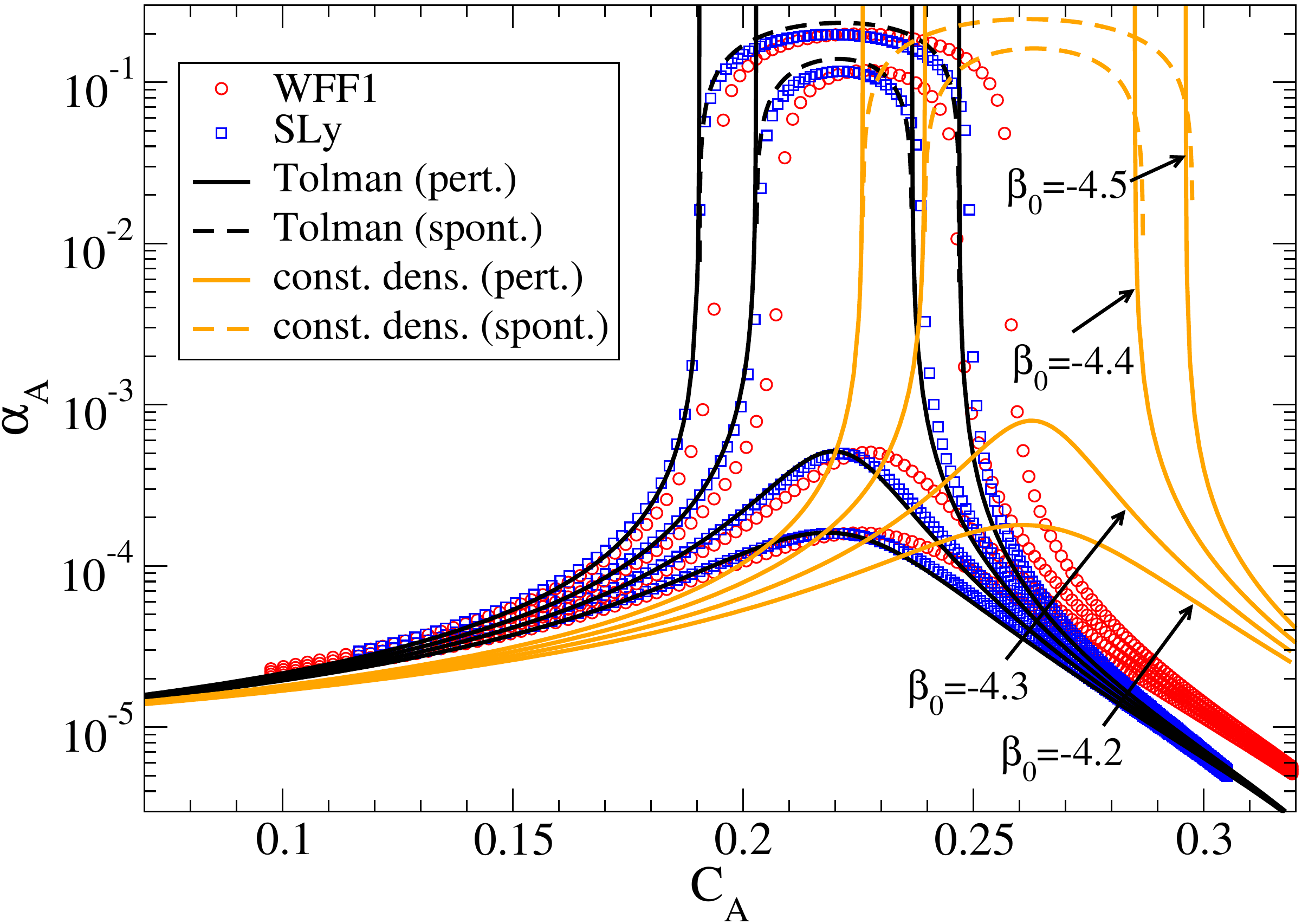}
\caption{\label{fig:charge-C-CD} Similar to Fig.~\ref{fig:charge-C-Tol} but including analytic results for constant density stars. Notice that scalar charges for such stars are quite different from those for realistic NSs. 
}
\end{figure}

% \begin{figure*}[!htbp]
% %\begin{center}
% \includegraphics[width=8.5cm]{charge-C-CD.pdf}
% \includegraphics[width=8.5cm]{charge-Bind-CD.pdf}\caption{\label{fig:charge-C-CD} (Left) Similar to Fig.~\ref{fig:charge-C-Tol} but including analytic results for constant density stars. Notice that scalar charges for such stars are quite different from those for realistic NSs. (Right) Similar to Fig.~\ref{fig:charge-Bind-Tol} but including analytic results for constant density stars. Unlike in the left panel, scalar charges for constant density stars are now similar to realistic NSs.
% }
% %\end{center}
% \end{figure*} 

Figure~\ref{fig:charge-C-CD} compares the analytic scalar charges for constant density stars (as a function of the compactness) with those for Tolman VII model and numerical charges with two representative EoSs. Notice that the spontaneous scalarization happens for larger compactnesses compared to the Tolman VII model and realistic NSs. This shows that the former is not an accurate model of the latter.

One can further convert the scalar charge expression in terms of the binding energy. For constant density stars, the leading relation is $\Omega_A/M_A = -(3/5) C_A + \mathcal{O}(\alpha_0^2)$.
% \begin{equation}
% \frac{\Omega_A}{M_A} = -\frac{3}{5} C_A + \mathcal{O}(\alpha_0^2)\,.
% \end{equation}
\begin{figure}[!htbp]
\includegraphics[width=8.5cm]{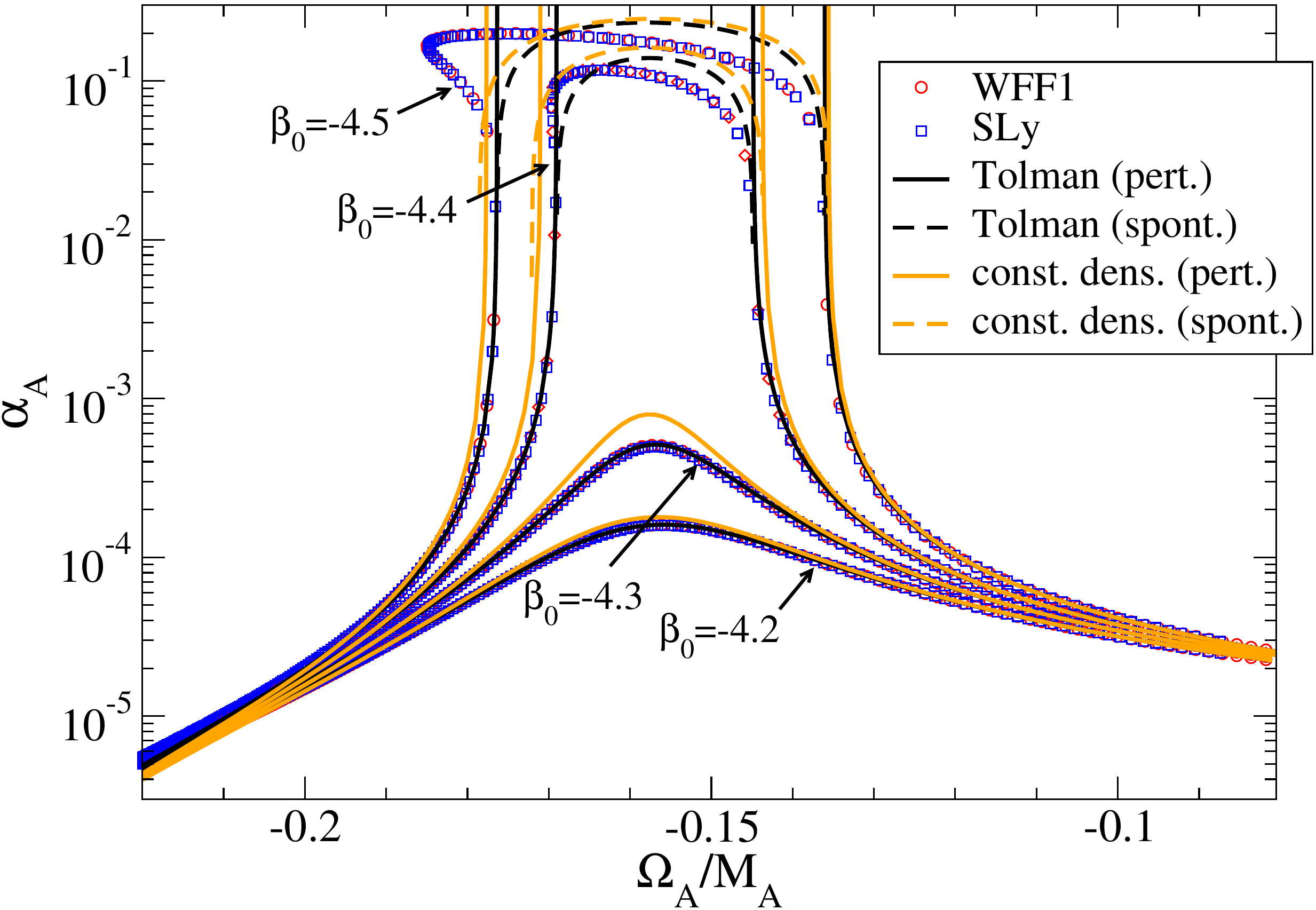}\caption{\label{fig:charge-Bind-CD} Similar to Fig.~\ref{fig:charge-Bind-Tol} but including analytic results for constant density stars. Unlike in Fig.~\ref{fig:charge-C-CD}, scalar charges for constant density stars are now similar to realistic NSs.
}
\end{figure}

Inverting this and substituting it into the analytic expression for the scalar charge in terms of the compactness, one finds the scalar charge as a function of the binding energy, which is shown in Fig.~\ref{fig:charge-Bind-CD}. We compare the constant density result with the Tolman VII one and numerical results with realistic EoSs. Observe that the scalar charges for constant density stars are very similar to other results, supporting the quasi-universality of the relation between the scalar charge and binding energy.

\bibliography{master}

\end{document}